\documentclass[%
 reprint,
showpacs,
 amsmath,amssymb,
 aps
]{revtex4-1}

\usepackage{graphicx}
\usepackage{dcolumn}
\usepackage{bm}
\usepackage{color}

\begin{document}

\preprint{APS/123-QED}

\title{Analysis of strong-field enhanced ionization of molecules using Bohmian trajectories}
\author {Ryohto Sawada$^{1,2}$}
\author{Takeshi Sato$^{2}$}
\author{Kenichi L. Ishikawa$^{1,2,3}$}
\affiliation{
$^{1}$Department of Applied Physics, Graduate School of Engineering, The University of Tokyo, 7-3-1 Hongo, Bunkyo-ku, Tokyo 113-8656, Japan\\
$^{2}$Photon Science Center, Graduate School of Engineering, The University of Tokyo, 7-3-1 Hongo, Bunkyo-ku, Tokyo 113-8656, Japan\\
$^{3}$ Department of Nuclear Engineering and Management, Graduate School of Engineering,The University of Tokyo, 7-3-1 Hongo, Bunkyo-ku, Tokyo 113-8656, Japan
}
\date{\today}

\begin{abstract}

We theoretically investigate the mechanism of enhanced ionization in two-electron molecules by analyzing Bohmian trajectories for a one-dimensional ${\rm H}_2$ in an intense laser field. We identify both types of ionizing trajectories corresponding to the ejection from the up-field and down-field cores. The { trajectories of the two electrons are correlated with each other in the former while correlation is negligible in the latter}. The contributions from the two ionization types, though depending on laser intensity and internuclear distance, are comparable to each other. 
\end{abstract}

\pacs{32.80.Rm, 33.80.Rv, 42.50.Hz}
\maketitle


\section{\label{sec:intro}Introduction}
\begin{figure}[t]
\includegraphics[width=8cm, bb = 0 0 565 232 ]{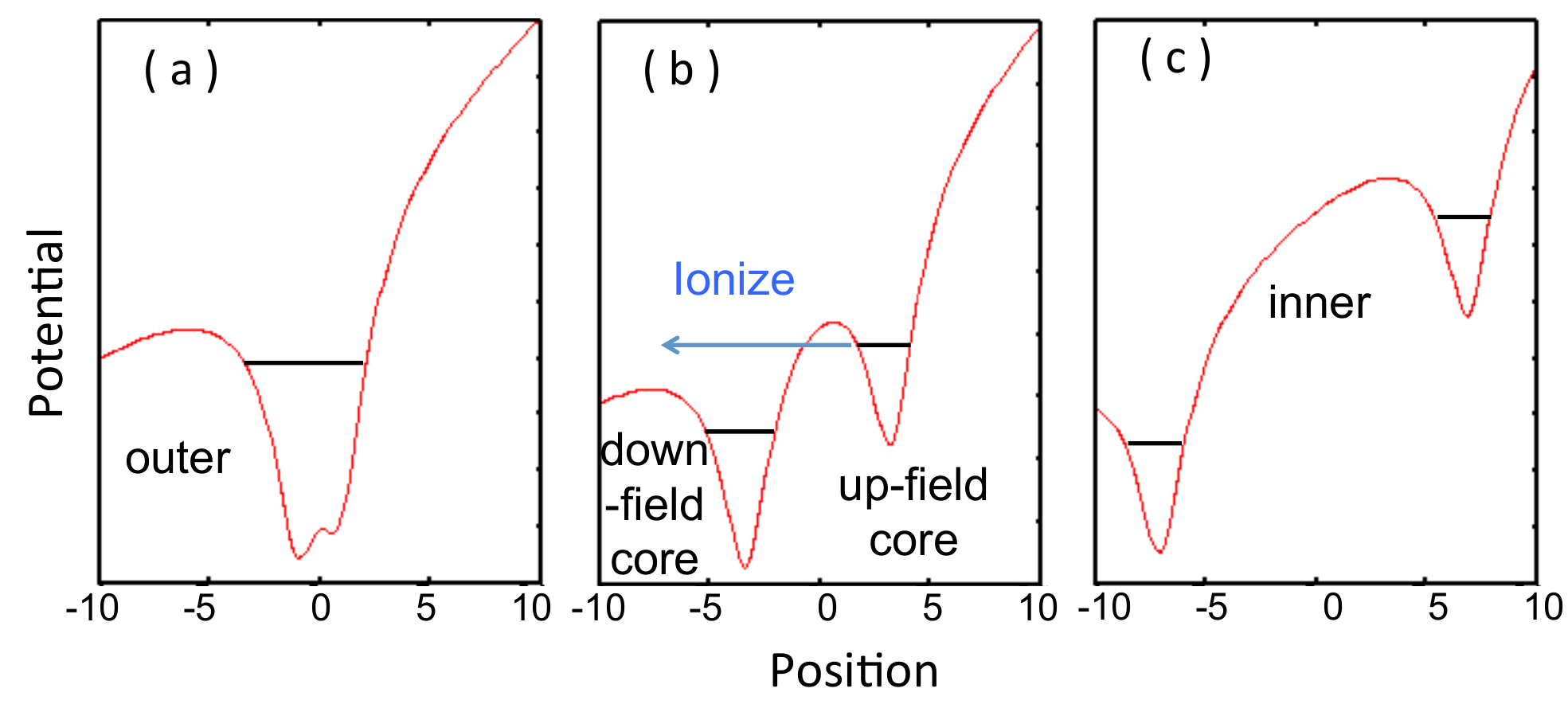}
\caption{{ } Schematic of the potential curves of a model diatomic molecule in a strong laser field at the (a) equilibrium, (b) critical, and (c) long internuclear distances.}
\label{fig:scheme}
\end{figure}

The ionization dynamics of atoms and molecules in intense femtosecond laser pulses is of great interest in a variety of fields such as attosecond science and coherent control of chemical reactions. It is widely accepted that the ionization probability of molecule depends on internuclear distance and strongly peaks at a certain critical distance longer than its equilibrium value \cite{band-enh-plo,cons-enh,mark-enhex,wei-enhex,Wu-enh}. This phenomenon, called enhanced ionization, is commonly believed to be due to a mechanism schematically depicted in Fig. \ref{fig:scheme} \cite{seide-enh}. At the equilibrium internuclear distance [Fig. \ref{fig:scheme} (a)], the electrons delocalize across the two cores and rarely tunnel through the outer potential barrier to the continuum. Around the critical distance, on the other hand [Fig. \ref{fig:scheme} (b)], the electrons begin to localize in each potential well, due to the rising inner barrier. Then, the electron in the up-field core can easily ionize through the inner barrier directly to the continuum. The ionization from the up-field core is expected to dominate that from the down-field core through the outer barrier, thus leading to a remarkable enhancement in the total tunneling ionization probability. Further increase in internuclear distance raises the inner barrier and hinders the inter-core tunneling, so the ionization probability decreases again [Fig.\ \ref{fig:scheme} (c)]. The genuineness of this view is, however, still under debate and a subject of active research\cite{Wu-enh,band-enh-lo,mulyukov-enh-lo,kawata-enh,harumiya-enh-3d}. {Wu {\it et al.} \cite{Wu-enh}, e.g., experimentally distinguished between the ejection from the up- and down-field cores in the ionization of ArXe by elliptically polarized pulses, and obtained the results that supported that from the up-field core.}

Various theoretical studies have been conducted to understand 
the enhanced ionization \cite{seide-enh,band-enh-lo,mulyukov-enh-lo,tkmt-enh,tkmt-enh-fix,tkmt-enh,tkmt-enh-fix,kawata-enh,harumiya-enh-3d,iskw-enh-pic,band-enh-plo}. In Refs.\ \cite{band-enh-lo,mulyukov-enh-lo,tkmt-enh,tkmt-enh-fix}, the role of the electron localization is examined using the time-dependent Schr\"odinger equation (TDSE) simulations for ${\rm H}_2^+$, either by projecting the wave function onto the localized state or by analyzing the temporal variation of the electron density distribution. It has been found that the transition rate from the state localized at the up-field core to the ionized state increases at the critical distance \cite{band-enh-lo,mulyukov-enh-lo}. On the other hand, simulations for one-dimensional { \cite{kawata-enh, hydro1d-mol}} and three-dimensional \cite{ale-enh-h2,harumiya-enh-3d} ${\rm H}_2$ have revealed that, around the critical internuclear distance, an ionic component ${\rm H}^+{\rm H}^-$ or ${\rm H}^-{\rm H}^+$ is created with a negatively-charged down-field core.
It is, however, not trivial to extract the information of ``from which part of the wave packet the ejected electron originates''. In this regard, particle-based approaches would be preferable. For example, particle-in-cell simulations of molecular multiple ionization \cite{iskw-enh-pic} have clearly identified test-particle trajectories that localize and ionize over the inner barrier directly to the continuum. Nevertheless, such a classical analysis cannot treat quantum-mechanical effects such as tunneling and the effect originating from the antisymmetry of the total wave function (including spin) under exchange of two electrons, referred to as effect of exchange hereafter.

\section{Bohmian trajectories}
\label{sec:bohmian-trajectories}

In this work, we address the mechanism of enhanced ionization using Bohmian trajectories \cite{bohm-honke-1,bohm-honke-2,sans-rev,det-bohm,orio-apl-bohm,quant-theo-mot,wyatt-book}  
{which can visualize the  probability current and have been applied to various fields such as strong field physics { \cite{song-bohm,takemoto-bohm,faria-hhg-2,faria-hhg-1}} and relativistic phenomena \cite{bens-rel}.} {The Bohmian trajectories are obtained by substituting the polar-form wave function 

\begin{eqnarray}
\label{eq:pol}
\psi(\vec{x},t) = A(\vec{x},t) \exp [iS(\vec{x},t)/\hbar]
\end{eqnarray}
to the TDSE, where $A$ and $S$ are real functions and $\vec{x} \equiv (x_{1},x_{2},...,x_{n})$ is the vector of electron positions. In this study, specifically, we consider a 1D ${\rm H}_2$ model molecule, thus we have $n=2$ and $\vec{x} = (x_{1},x_{2})$.
{ Then, we obtain the equation of continuity:
\begin{equation}
\label{eq:cont}
\frac{\partial A^{2}(x_{1},x_{2},t)}{\partial t} + \sum^{2}_{i=1} \frac{\partial}{\partial x_{i}} \left [  j_{i}(x_{1},x_{2},t)  \right ]=0,
\end{equation}
where, 
\begin{equation}
\label{eq:curr}
j_{i} \equiv A^{2}\frac{\partial S}{m\partial x_{i}}=i\frac{\hbar}{2m}(\psi \frac{\partial \psi^{*}}{\partial x_{i}} - \frac{\partial \psi}{\partial x_{i}}  \psi^{*}),
\end{equation}
denotes the particle current density, and the quantum Hamilton-Jacobi equation:}
\begin{eqnarray}
\label{eq:phase}
\frac{\partial S(x_{1},x_{2},t)}{\partial t} &&+ \sum^{2}_{i=1} \left[  \frac{1}{2m} \left(\frac{\partial S(x_{1},x_{2},t)}{\partial x_{i}} \right)^{2} \right] \nonumber \\
&&+ U(x_{1},x_{2},t) + Q(x_{1},x_{2},t) =0,
\end{eqnarray}
where $U(x_{1},x_{2},t)$ denotes the classical potential. Equation (\ref{eq:phase}) differs from the classical counterpart only by the quantum potential:

\begin{eqnarray}
\label{eq:qpot}
Q(x_{1},x_{2},t) = \sum^{2}_{i=1} \frac{-\hbar^{2}}{2m}\frac{\partial^{2}A(x_{1},x_{2},t)/\partial x^{2}_{i}}{A(x_{1},x_{2},t)},
\end{eqnarray}
which describes the quantum-mechanical effects beyond the classical dynamics. 

Let us regard the probability density as a fluid and consider a test particle co-moving with the probability flow with a
velocity,
{
\begin{eqnarray}
\label{eq:vel}
v_{i} \equiv \frac{j_i}{|\psi|^2}=\frac{j_i}{A^2}=\frac{\partial S}{m\partial x_{i}}=\frac{1}{m}\Re \left( \frac{-i\hbar\,\partial \psi / \partial x_i}{\psi}\right).
\end{eqnarray}
}
%
Then, the path of a fluid particle coincides with a Bohmian trajectory (see, e.g., \S 3.10 of \cite{quant-theo-mot}) satisfying the following equation of motion:
{ 
\begin{eqnarray}
\label{eq:dyna}
{m\frac{d v_{i}(t)}{d t}} &=& {m\frac{d^2 x_{i}(t)}{d t^2}} \nonumber \\
 &=& -\frac{\partial}{\partial x_{i}} \left[ U(x_{1},x_{2},t) + Q(x_{1},x_{2},t)  \right ].
\end{eqnarray}
}
{Thus, we can use the Bohmian trajectories as tracer particles, with which we can clearly visualize the flow of the electron wave packet, by analogy to classical flow visualization with, e.g., fluorescent beads}. If the initial position is distributed according to $|\psi(\vec{x},t=0)|^2$ with vanishing velocity, a whole set of trajectories convey the information completely equivalent to the wave function 
{\cite{det-bohm,orio-apl-bohm,quant-theo-mot,wyatt-book}}. 

{Let us briefly list features of Bohmian trajectories relevant to the present study. For more details, we refer the reader to literature \cite{det-bohm,orio-apl-bohm,quant-theo-mot,wyatt-book}

\begin{itemize}

\item The quantum potential introduces all quantum effects including tunneling, nonlocality, and the effect of exchange.

\item As is clear from Eq.\ (\ref{eq:curr}), Bohmian particles do not move, i.e., $v_i=0$ for the case of a stationary state whose wave function is real except for a global phase factor. The classical force (Coulomb force from the nucleus and the other electron) is cancelled by the force derived from the quantum potential.

\item Bohmian trajectories can pass through a classically forbidden spatial region due to the quantum potential $Q$, capable of describing tunneling.

\item The quantum potential depends on $A(x_1,x_2,t)$, i.e., the distribution of trajectories. Therefore, different trajectories $(x_1(t), x_2(t))$ and $(x^\prime_1 (t), x^\prime_2 (t))$ influence each other, unlike in the classical mechanics where they can be calculated independently. { This is true for a single-electron case as well; different trajectories $x(t)$ and $x^\prime (t)$ influence each other through the quantum potential. A familiar consequence is the interference fringe in the double-slit experiment (p.\,53 of Ref.\ \cite{orio-apl-bohm}).}

\item Trajectories { $x_1(t)$ and $x_2(t)$} of two identical particles cannot cross each other, which is a direct consequence of the (anti-)symmetrization of the total wave function. For example, the quantum potential exerts a repulsive force to prevent the crossing of trajectories of two counter-propagating free identical particles {\it even without any classical interaction}  (see Appendix).

\end{itemize}
}

{ Interpretations of high-field phenomena drawn from Bohmian trajectories \cite{song-bohm,takemoto-bohm,faria-hhg-2,faria-hhg-1} may be different from those based on other approaches, e.g., Feynman's path-integral \cite{sall-sci-path} and quantum-orbit approaches \cite{Lewenstein1994PRA,mil-pra-orbit}.}

Although Bohmian mechanics (or de Broglie-Bohm theory), which describes quantum systems in terms of Bohmian trajectories, is often studied in the context of the interpretation of quantum mechanics, we stress that we are not concerned with this conceptual aspect in the present study. We use Bohmian trajectories exclusively as tracer particles.
Our analysis identifies two classes of ionizing trajectories around the critical internuclear distance. The one corresponds to the ejection from the up-field core in which the two electrons move in a concerted way, while the other to the ejection from the down-field core.

\section{Simulation}

We consider the one-dimensional model for H$_{2}$ in which two protons are fixed on the $x$-axis and the electronic motion is restricted along this axis. {Although the inclusion of the nuclear quantum dynamics 
would be ideal, it would be very time-consuming. Moreover,} it is expected that the dynamics of the protons have little effect on those of electrons \cite{tkmt-enh-fix}. The field-free Hamiltonian is given as (atomic units are used throughout):
\begin{eqnarray}
\label{eq:ham}
\scalebox{0.8}{$\displaystyle
H= -\frac{1}{2}\sum^{2}_{i=1} \left[ \frac{d^{2}}{dx_{i}^{2}} - \frac{1}{\sqrt{ \left ( x_{i}\pm \frac{R}{2} \right)^{2} +\alpha }} \right ] + \frac{1}{\sqrt{ \left ( x_{1}-x_{2} \right)^{2} +p }}
$}
\end{eqnarray} 

where $R$ denotes the internuclear distance, and $p$ = 1.2375 and $\alpha$= 0.7 are the soft Coulomb parameters \cite{band-h2}. The laser wavelength and intensity are assumed to be 1064 nm and $1.7 \times 10^{14}$ W/cm$^{2}$ with a 5-cycle ramp-up, respectively [Fig.\ \ref{fig:types} (a)]. 

We numerically solve the TDSE {in the length gauge} by basis set approach in the constrained interpolation profile method (CIP-BS$^{1}$) \cite{utm-cip}. The Bohmian trajectories are obtained by numerically integrating Eq. (\ref{eq:dyna}) with the quantum potentials $Q$ derived from the instantaneous wave function using Eq. (\ref{eq:qpot}). Initial positions $(X_{1},X_{2})= (x_1(t=0),x_2(t=0))$ are sampled according to the ground state probability density, with vanishing initial velocities.

{ 
\section{Results and Discussions}
}

In the blue { cross} in Fig.\ \ref{fig:prob} (a), we show the ionization probability $P$ as a function of internuclear distance $R$, calculated as $P=1-\int\int_{\mid x_{1}\mid, \mid x_{2}\mid <15}  |\psi|^{2} dx_1dx_2$ at $t=1000$. 
When $R$ increases from its equilibrium value $R_{e}=$1.68, $P$ increases, peaks at the critical distance $R_{c}$ = 5.6, and decreases again.

\begin{figure}[t]
\includegraphics[width=8cm, bb = 0 0 527 558]{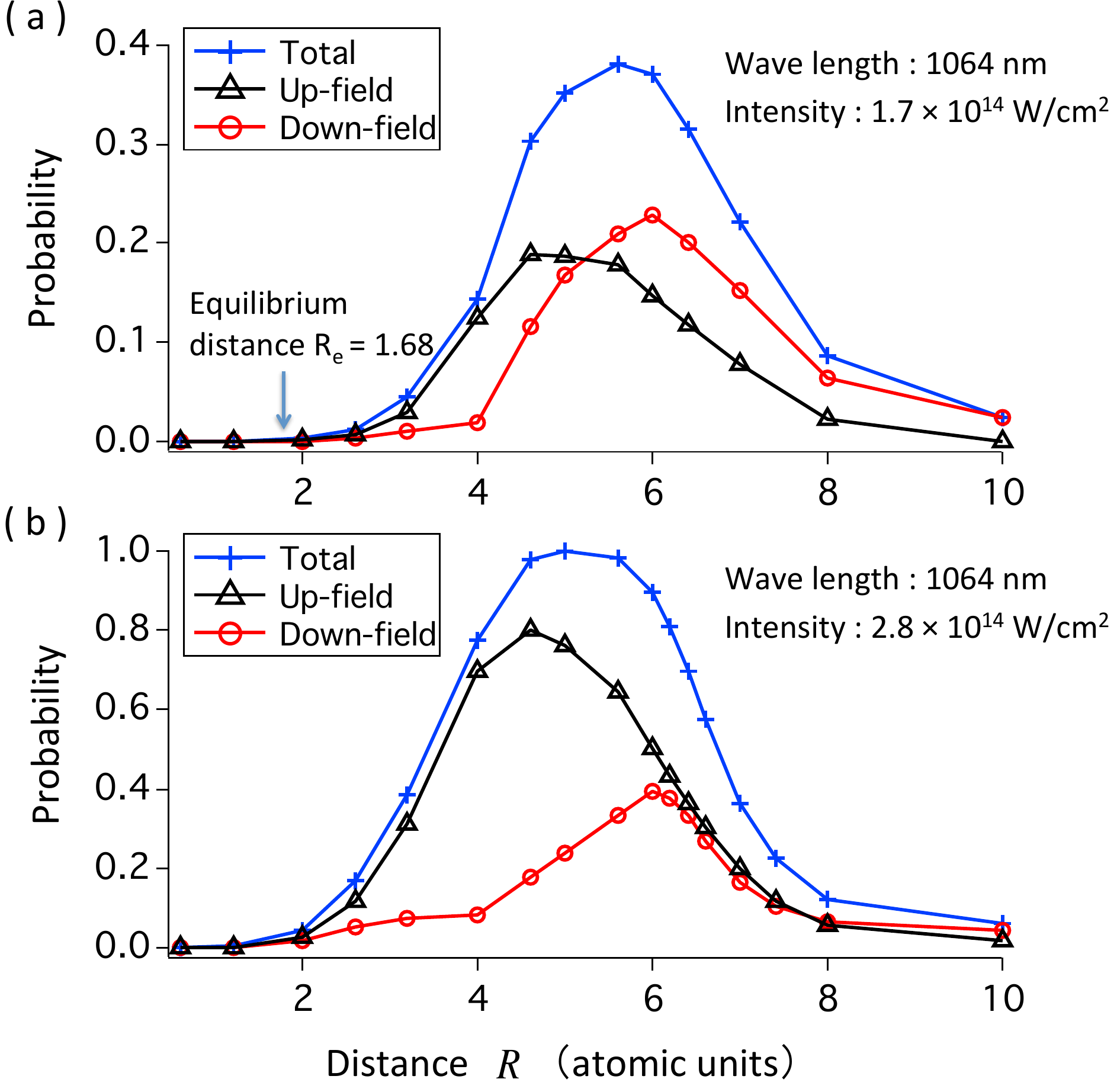}
\caption{{ } {Total ({blue cross}), Type 1 ({black triangle}), and Type 2 ({ red circle}) ionization probabilities in different laser intensities (a) $1.7 \times 10^{14}$ W/cm$^{2}$ and (b) $2.8 \times 10^{14}$ W/cm$^{2}$ with a 5-cycle ramp-up.}}
\label{fig:prob}
\end{figure}

\begin{figure*}[t]
\includegraphics[width=16cm, bb = 0 0 679 254]{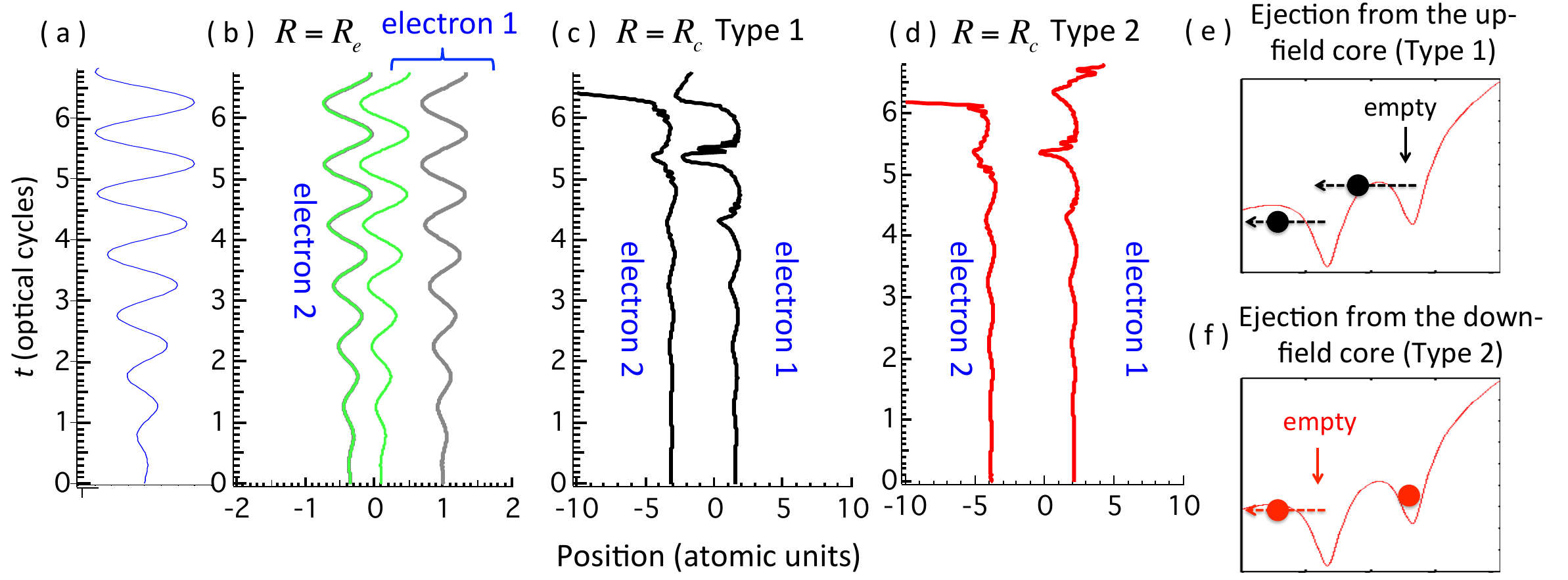}
\caption{{ } Bohmian trajectories for different combinations of the initial positions $(X_{1},X_{2})$ and internuclear distances $R$. (a) Laser electric field. (b) The trajectories at the equilibrium distance for the case of $(X_{1},X_{2}) =$  (1,-0.35) {(gray)} and (0.1,-0.35) {(green)}. The two lines for electron 2 overlap each other. (c) The trajectory at the critical distance for the case of $(X_{1},X_{2}) =$  (1.5,-3.15) (Type 1). (d) The trajectory at the critical distance for the case of $(X_{1},X_{2})=$ (2.1,-3.85) (Type 2). { The trajectories in (c) and (d) are shown on the $(x_1,x_2)$ plane in Fig.\ \ref{fig:exflu} (b) and (a), respectively.} (e,f) Schematic of the ejection from the (e) up-field core and (f) the down-field core.}\label{fig:types}
\end{figure*}

In Fig.\ \ref{fig:types} (b), (c) and (d), we plot the Bohmian trajectories $(x_{1}(t),x_{2}(t))$ for different combinations of the initial positions and internuclear distances. In the equilibrium distance [Fig.\ \ref{fig:types} (b)], the electrons oscillate around their initial positions, closely following the laser field. For $(X_{1},X_{2})=(0.1,-0.35)$, electron 1 oscillates between the two potential wells, undisturbed by the inner barrier. We also notice that the trajectory of electron 2 changes little when the initial position of electron 1 is changed (compare {gray(dark) and green(light)} lines), indicating that the electronic correlation is negligible.

We now focus on the ejected trajectories at the critical distance. 
The ejected trajectories are classified into two types, whose typical examples are plotted in Fig.\ \ref{fig:types} (c) and (d). In Type 1 [Fig.\ \ref{fig:types} (c)], electrons 1 and 2 are initially localized in the right and left core, respectively. At $t \approx 5\frac{1}{4}T$, with $T$ being the optical cycle, electron 1 once tunnels through the inner barrier but returns to the right {presumably through tunneling}. One optical cycle later, finally, it tunnels to the left (down-field) well while, at the same time, electron 2 is ejected to the left. 

{ 
At first sight, it would appear that electron 1 dynamically kicks electron 2 out through the interelectronic Coulomb repulsion. However, also 
the effect of exchange exerted through the quantum potential plays an important role in the ejection of electron 2 instead of electron 1. As a remarkable manifestation of the effects of identical particles in Bohmian mechanics (see, e.g., \S 7.1.4 of \cite{quant-theo-mot} for the discussion on identical particles in Bohmian mechanics), one can indeed demonstrate that, for the antisymmetrized two-electron wave function, the trajectories of the two electrons do not cross each other, and that the apparent interelectronic kick would be seen \emph{even without any classical interaction between the two electrons} (see Appendix and Fig.\ \ref{fig:nc}). The quantitative comparison of the contributions from the Coulomb repulsion and the exchange is difficult, since the quantum potential involves not only the effect of exchange but also other non-classical effects such as tunneling. We have found, however, that not electron 1 but 2 is ejected even if we switch off the interelectronic soft Coulomb potential [the second term of Eq.\ (\ref{eq:ham})] at $t>6.2T$ in the simulation { and, on the other hand, that electron 2 passes by electron 1 and escapes from the molecule if we switch off the quantum potential at $t>6.2T$ while retaining the interelectronic repulsion (see Fig.\ \ref{fig:aptr}). These observations imply} that the apparent kick-out in Fig.\ \ref{fig:types} (c) is probably mainly due to the effect of exchange, rather than a dynamical process due to the Coulomb repulsion. From these considerations, we may conclude that Type 1 corresponds to \emph{ionization from the up-field core} in the context of the discussion on the mechanism of enhanced ionization.}  

In Type 2 [Fig.\ \ref{fig:types}  (d)], on the other hand, electron 1 is basically localized in the right core all the time. This type, hence, { is identified as} {\it ionization from the down-field core} [Fig. \ref{fig:types} (f)].

{ We have also analyzed a higher intensity case ($2.8\times 10^{14}\,{\rm W/cm^2}$) and found that ionizing trajectories (not shown) can be classified into types that have characteristics similar to those in Figs.\ \ref{fig:types}  (c) and (d) { while electrons are more strongly shaken and ionization takes place earlier.}}

\begin{figure}[b!]
\includegraphics[width=8cm, bb = 0 0 538 418]{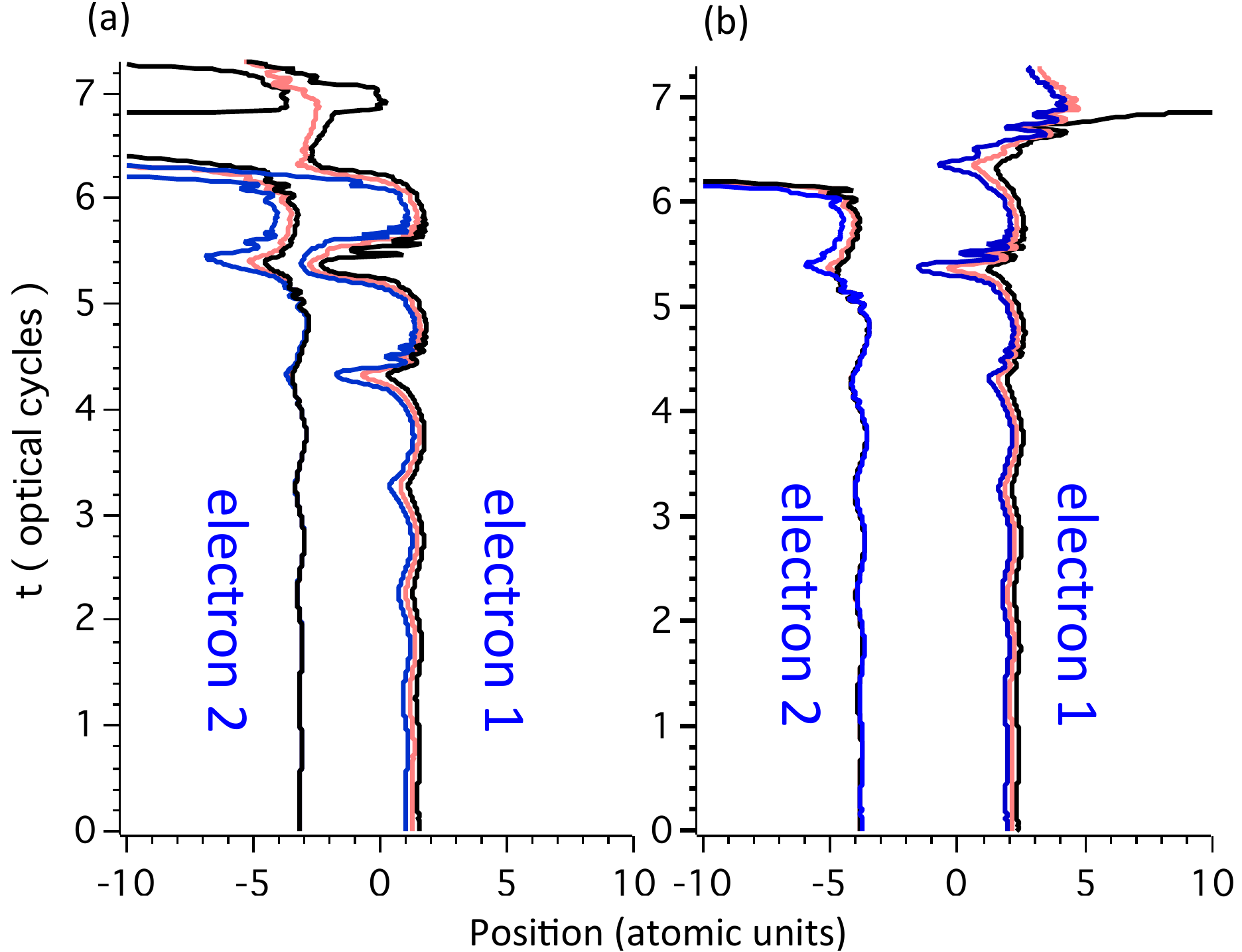}
\caption{{ } Bohmian trajectories at $R=R_{c}$ till $t=7.25T$. (a) Comparison of the trajectories for the case of ($X_{1},X_{2}$) = (1.5, -3.15)(black), (1.25, -3.15)({ pink}) and (1.0,-3.15)({ blue}) (Type 1 ionization). (b) Comparison of the trajectories for the case of ($X_{1},X_{2}$) = (2.35, -3.85)(black), (2.1, -3.85)({ pink}) and (1.9,-3.85)({ blue }) (Type 2 ionization).}
\label{fig:after}
\end{figure}

In Fig\ \ref{fig:after} we examine the dynamics of the partner electron [electron 1 in the case of Fig.\ \ref{fig:types}(c,d)] after the ionization, which are found to vary depending on its initial position. For the case of Type 1 ionization [Fig\ \ref{fig:after}(a)], in particular, the trajectory of electron 1 is much more sensitive to its initial position after the ejection of electron 2 than before. It is interesting to notice that black lines correspond to rescattering of electron 1. For the case of Type 2 ionization [Fig\ \ref{fig:after}(b)], the motion of electron 1 looks less stable after the ejection of election 1, possibly due to sudden change in the potential experienced by electron 1 and also because the remaining ion is left in a superposition of different states, still shaken by a strong laser field { (Unfortunately, the Bohmian trajectory analysis does not provide direct information on a superposition in which the ion is produced).} The computational cost would become increasingly higher for longer simulations, since basically we have to keep both electrons within the simulation box for the purpose of this study. Complete discussion on the partner electron dynamics after ionization, affected also by the pulse shape and molecular dissociation, as well as the Bohmian trajectory analysis of rescattering-induced phenomena such as non-sequential double ionization would require further investigation. Anyway, since the motion of electron 1 after the ejection of electron 2 does not influence the latter, we can discuss the mechanism of enhanced ionization independently from the details of the partner electron dynamics after it. { From an experimental viewpoint, besides the method in Ref.\ \cite{Wu-enh}, the wave packet created at the time of ionization may be accessed by high-harmonic spectroscopy \cite{nat-hhgspectra,hae-nat-hhgspectra,worner-prl-hhgspectra,hydro1d-mol}}  

\begin{figure*}
\includegraphics[width=16cm, bb = 0 0 697 261]{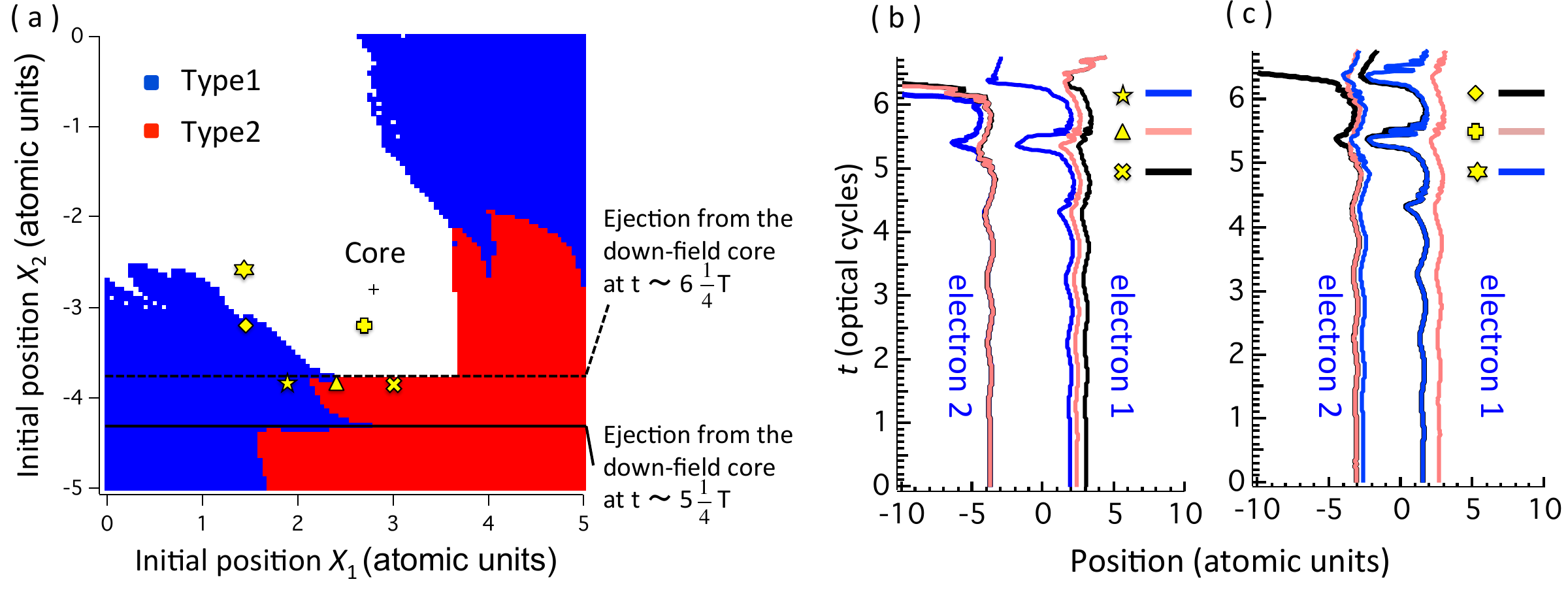}
\caption{{ } (a) Area of initial positions leading to the Type 1 and 2 ionization. Black dashed line shows the boundary between Type 2 and no-ionization areas, and black solid line the boundary between Type 2 ionization during the sixth optical cycle and that during the seventh cycle. (b) Comparison of the trajectories for the case of $(X_{1},X_{2})=(3,-3.8)$ (black), (2.35,-3.8) ({pink}) and (1.85,-3.8) ({ blue}). (c) Comparison of the trajectories for the case of $(X_{1},X_{2})=(1.5,-3.15)$ (black), (2.65,-3.15) ({ pink}) and (1.5, -2.65) ({ blue}).}
\label{fig:map}
\end{figure*}

The role of the electronic correlation, of both classical and quantum origins, in the ionization process becomes clearer, if we map the area of initial positions leading to the Type 1 and 2 ionization, respectively, at the critical distance [Fig. \ref{fig:map} (a)]. { Here, for convenience, we judge the type of ionization based on the potential well to which the partner electron belongs to at the moment of ionization. For example, if $x_1<0$ ($>0$) when electron 2 is ejected to the left, the ionization is classified as Type 1 (Type 2). While one cannot unambiguously define the time of ionization, again for convenience, we consider that an electron is ejected once its distance from the origin exceeds 15 a.u.} 

{ In Fig. \ref{fig:map} (a),} we can see that the boundaries between Type 2 and no-ionization areas are parallel to the horizontal or vertical axis (black dashed line). More detailed analysis shows that the boundaries between Type 2 ionization during the sixth optical cycle and that during the seventh cycle are also parallel to the axes (black solid line). This observation is indicative of a minor role played by the electronic correlation in Type 2 ionization (ionization from the down-field core). Indeed, the comparison of the { pink (light gray) and black lines} in Fig.\ \ref{fig:map}(b) reveals that the trajectory of the ejected electron hardly depends on that of the partner electron.

However, their behavior changes as the initial position of the partner electron comes closer to zero and $(X_{1},X_{2})$ enters the Type 1 area in Fig. \ref{fig:map} (a). In the blue curve in Fig.\ \ref{fig:map} (b) and the black curve in Fig.\ \ref{fig:map} (c), electron 1 can tunnel through the inner potential barrier and migrate to the down-field core at $t \approx 5\frac{1}{4}T$ and $t \approx 6\frac{1}{4}T$. This influences the trajectory of electron 2, in contrast to the situations at $t \lesssim 5T$, for the Type 2 ionization [Fig.\ \ref{fig:map} (b)], and for the equilibrium distance [Fig.\ \ref{fig:types} (b)] in which the trajectory of electron 2 is basically unaffected by that of electron 1. The structure of the boundary between the Type 1 ionization and no-ionization and that between the Type 1 and 2 ionization areas, neither horizontal nor vertical, reflects { the fact that the trajectories of the two electrons are correlated with each other in Type 1 (ionization from the up-field core).}

Although the ionization from the up-field core has been often highlighted as a mechanism of the enhanced ionization, we find that the ejection from the down-field core contributes significantly as well. In Fig. \ref{fig:prob} we plot the ionization probability of each type as a function of internuclear distance, {evaluated} by counting corresponding trajectories weighted by the initial probability distribution $|\psi (x_1,x_2,t=0)|^2$. Whereas their relative contribution varies with the internuclear distance and laser intensity, both types of ejection are equally important, overall.

{ Then, why is the ejection from the down-field core also enhanced? 
This is, unfortunately, difficult to understand just by looking at a potential curve as in Fig.\ \ref{fig:scheme}, or based on Bohmian trajectories and their equation of motion Eq.\ (\ref{eq:dyna}) alone.
In order to get an insight, instead, let us plot in Fig.\ \ref{fig:exflu} the snapshots of the electron probability density $|\psi|^2$ and its current density $\vec{j}$ (or equivalently the density and flux of Bohmian particles), calculated using the instantaneous wave function obtained from the TDSE simulations. { In order to visualize how Bohmian particles move on the $(x_1,x_2)$ plane, we also depict the ionizing trajectories in Fig.\ \ref{fig:types} (c) and (d) as red {(gray)} lines in Fig.\ \ref{fig:exflu} (b) and (a), respectively.}
Note that the laser electric force acting on the electrons points to the negative direction. In Fig.\ \ref{fig:exflu} (a) for $R=6.4$ and $t = 6.1 T$, we see some ionic component { (part of the wave function at $x_1,x_2>0$ or $x_1,x_2<0$)} ${\rm H}^+{\rm H}^-$  with a negatively charged up-field core (around $x_1=x_2=3.2$, marked as A) formed due to the rising inner barrier \cite{iskw-enh-pic}. At the same time, we also see other small components marked as B and C flowing to the negative direction, which correspond to Type 2 ionization. The ionic component A is not long-lived but, at slightly later time $t=6.27T$ where the laser field is stronger, it moves to the down-field atom [Fig.\ \ref{fig:exflu} (b)] to form an ionic component ${\rm H}^-{\rm H}^+$ { \cite{kawata-enh,hydro1d-mol}} with a negatively charged down-field core, marked as D. A part of it flows even further to the negative direction (marked as E and F), which corresponds to Type 1 ionization { (see also Fig.\ 1(b) of \cite{kawata-enh} and Fig.\ 2 of \cite{hydro1d-mol})}. Hence, Type 1 ionization takes place slightly later than Type 2 within each half cycle, as can be confirmed in Figs.\ \ref{fig:types} and \ref{fig:after}. At $R=2$ [Fig.\ \ref{fig:exflu} (c)] and 10 [(d)], on the other hand, the probability density distribution on the $(x_1,x_2)$ plane is much less deformed than at the critical distance, and neither ionic components nor ionizing currents can be seen in the scale of the figure.

\begin{figure}
\includegraphics[width=8cm, bb = 0 0 520 520]{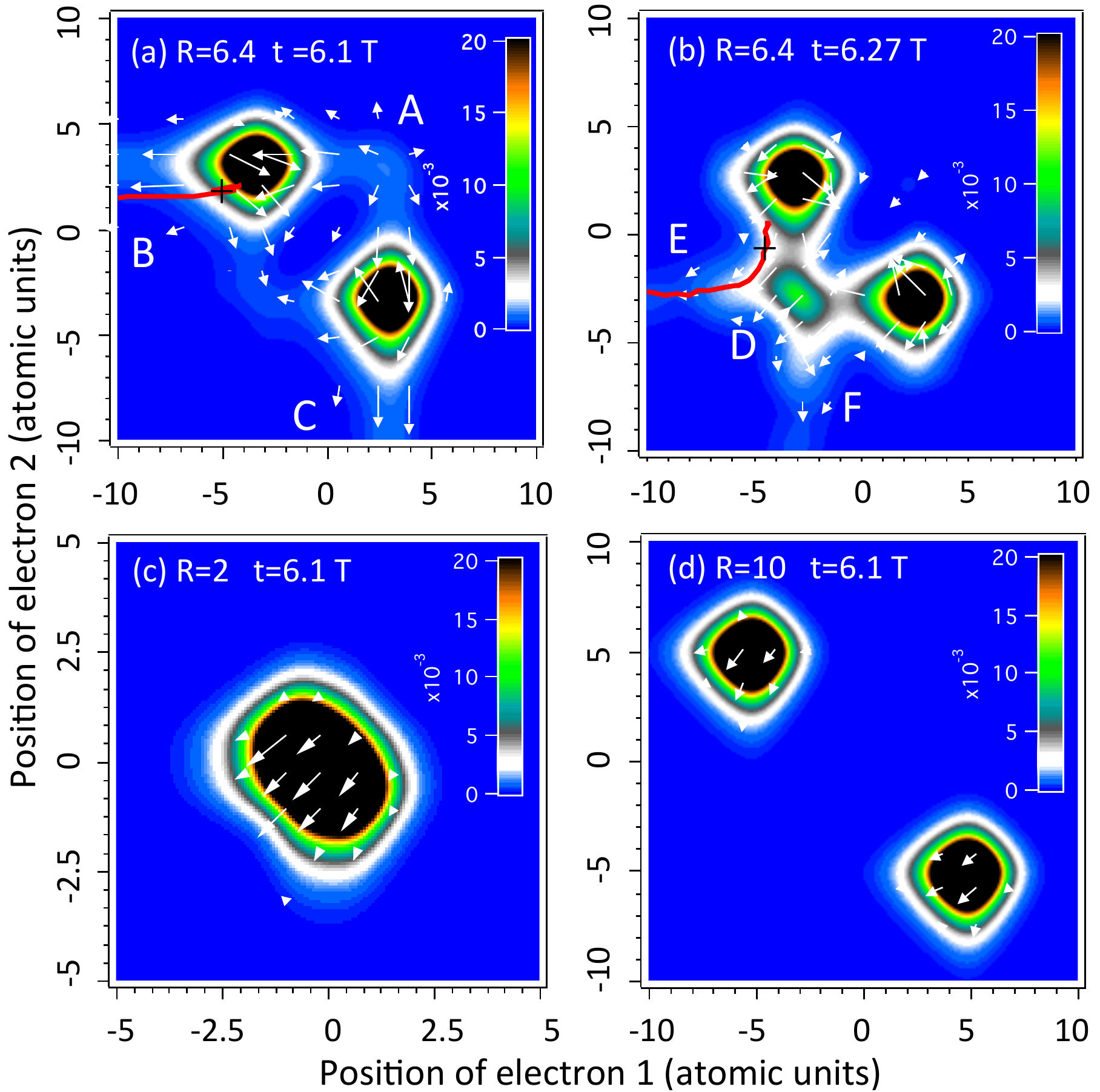}
\caption{{{ }  Snapshots of the electron probability density $|\psi (x_1,x_2)|^2$ (false color plot) and the particle current density $(j_1,j_2)$ (arrow), obtained from the TDSE simulations, at (a) $R=6.4$ a.u., $t = 6.1\, T$, (b) $R=6.4$ a.u., $t = 6.27\, T$, (c) $R=2.0$ a.u., $t = 6.1\, T$, and (d) $R=10$ a.u., $t = 6.1\, T$. The electric field (force acting on the electrons) is in the positive (negative) direction.} { Red(gray) lines on panels (a) and (b) depict the trajectories in Fig.\ \ref{fig:types} (d) and (c), respectively, with black crosses indicating the positions at  $t = 6.1\, T$ and $6.27\, T$, respectively.}}
\label{fig:exflu}
\end{figure}

\begin{figure}
\includegraphics[width=8cm, bb = 0 0 512 530]{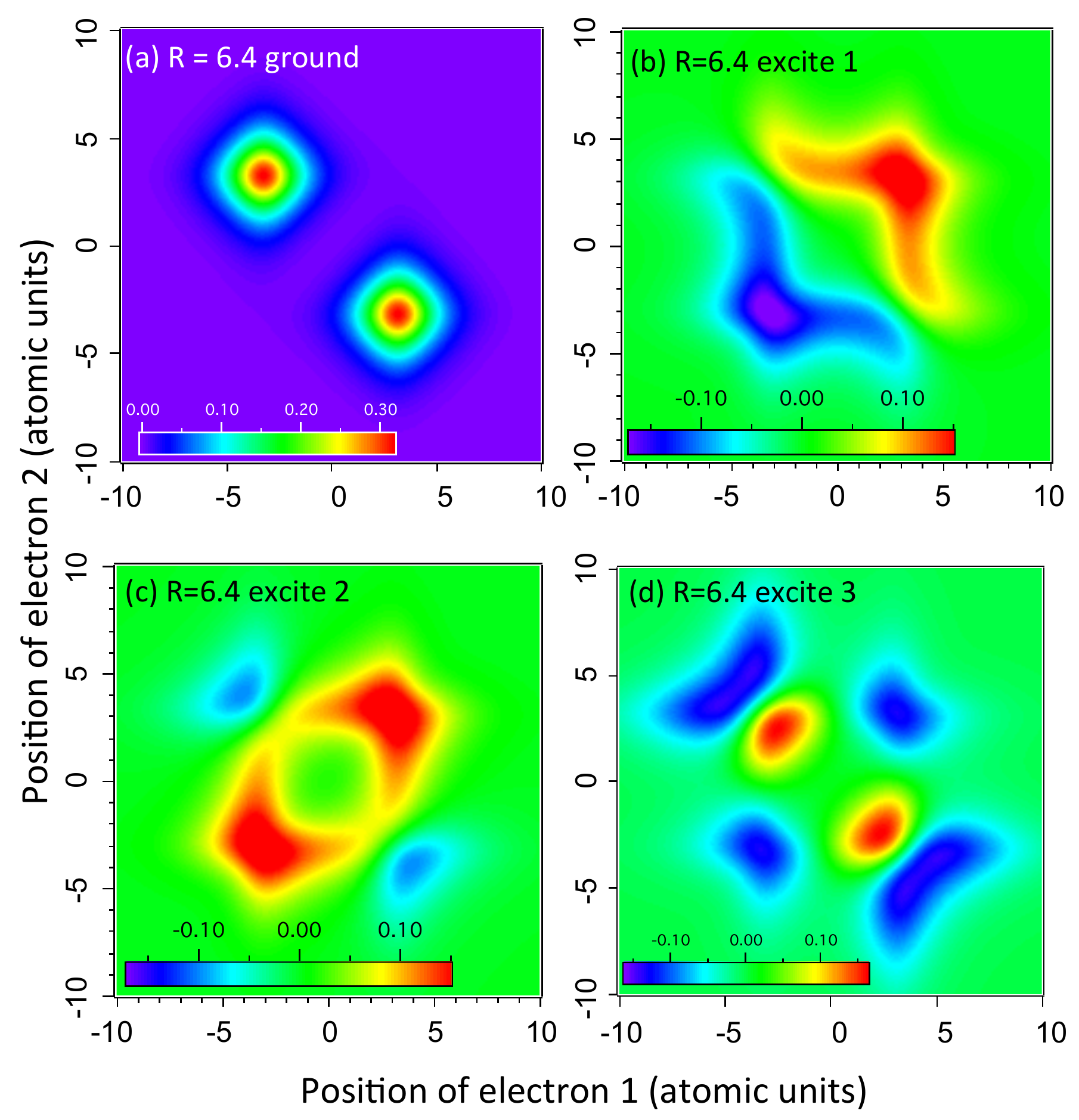}
\caption{{{ } Wave function of the { field-free} (a) ground, (b) first, (c) second and (d) third excited states at $R=6.4.$}}
\label{fig:excites}
\end{figure}

The wave function $\psi (x_1,x_2,t)$, at any moment, can be expressed as a superposition of field-free eigenstates. The amplitudes and phases of the (field-free) ground and first three excited states, whose wave functions at $R=6.4$ are shown in Fig.\ \ref{fig:excites}, are listed in Table\ \ref{ta:coeff} for each panel in Fig.\ \ref{fig:exflu}. At $R=2.0$, the wave function is mainly composed of the ground and first excited states, while it is dominated by the ground state at large distance ($R=10.0$). At the critical internuclear distance ($R=6.4$), in contrast, we find participation from many excited states; indeed, we have to take more excited as well as continuum states into account in order to achieve convergence. Such excitation taking place preferentially around the critical distance is consistent with previous reports \cite{kawata-enh}.} From the shapes of eigenstates shown in Fig.\ \ref{fig:excites}, one may be reasoned that when the ionic component A in Fig.\ \ref{fig:exflu} (a) is formed in the up-field core, components B and C undergoing Type 2 ionization can also be formed at the same time; the former itself can be intuitively understood as an effect of the rising inner barrier. On the other hand, for larger internuclear distance, dominated by the ground state, the formation of such components are unlikely. { Hence, Type 2 ionization as well as Type 1 is enhanced around the critical internuclear distance. In terms of Bohmian mechanics, the co-occurrence of the component A and the components B, C is presumably due to the mutual influence between the trajectories forming the former and those forming the latter (see the fourth item on the list in Sec.\ \ref{sec:bohmian-trajectories}), which is, unfortunately, hard to grasp intuitively.}

We see from Fig.\ \ref{fig:prob} that the ratio of Type 2 ionization to Type 1 increases with $R$. This can be qualitatively understood as follows. With increasing internuclear distance, the rising inner barrier leads to more formation of the delayed ionic component with a negatively charged up-field core, and, at the same time as discussed above, to more ejection form the down-field core.

\begin{table}
 \begin{tabular}{l c c c c}\hline\hline
 & (a) & (b) & (c) & (d) \\ \hline
ground state, amplitude & 0.866 & 0.829 & 0.98 &0.99 \\ 
ground state, phase & 2.36 & 0.95 & -0.16 &3.17 \\ 
1st excited state, amplitude & 0.132 & 0.09 & 0.145 & 0.001 \\ 
1st excited state, phase & -0.06 & 2.04 & 3.09 & 2.86 \\ 
2nd excited state, amplitude & 0.166 & 0.20 & 0.01 & $<$10$^{-8}$ \\ 
2nd excited state, phase & -0.02 & -1.66 & 3.42 & -- \\ 
3rd excited state, amplitude & 0.077 & 0.035& $<$10$^{-4}$ &0.001 \\ 
3rd excited state, phase & 1.30 & 0.332&  -- & 4.02 \\ \hline\hline
\end{tabular}
\caption{\label{ta:coeff} {{ The amplitudes and phases (in radian) of the field-free ground and the first three excited states (Fig.\ \ref{fig:excites}) contained in each instantaneous wave function shown in Fig.\ \ref{fig:exflu}. (a) $R=6.4$ a.u., $t = 6.1\, T$, (b) $R=6.4$ a.u., $t = 6.27\, T$, (c) $R=2.0$ a.u., $t = 6.1\, T$, (d) $R=10$ a.u., $t = 6.1\, T$}}}
\end{table}

\section{\label{sec:conc}Conclusion}

We have investigated enhanced ionization of a 1D H$_{2}$ model molecule using Bohmian trajectories. The Bohmian trajectories { can} visualize { flow} of the probability density and illustrate how ionization proceeds. Around the critical internuclear distance, not only the ejection from the up-field core, involving tunneling through the inner barrier, but also that from the down-field core through the outer barrier are comparably enhanced, although their relative contribution varies with the internuclear distance and laser intensity. Our analysis has also revealed that the trajectories of the two electrons are correlated with each other for the case of the ejection from the up-field core, while { the trajectory of} the departing electron is not affected by { that of} the other electron for the case of the ejection from the down-field core.

\begin{acknowledgments}
This research is supported in part by Grant-in-Aid for Scientific
 Research (No. 23750007, No. 23656043, { No. 25286064, and No. 26600111}) from the
 Ministry of Education, Culture, Sports, Science and Technology (MEXT)
 of Japan, and also by the Photon Frontier Network program of MEXT, Japan { and JSPS KAKENHI (26-10100)}. { This research is also partially supported by the Center of Innovation Program from Japan Science and Technology Agency, JST.} R. S. gratefully acknowledges support by Graduate School of Engineering, the University of Tokyo Doctoral Student Special Incentives Program (SEUT Fellowship).
\end{acknowledgments}

\appendix

{ \section{Non-crossing of the Bohmian trajectories of two electrons \label{sec:appnc}}
 
In the spin singlet state, as is considered in the present study, the spatial wave function $\psi (x_1,x_2,t)$ is symmetric under the exchange of the two electrons, i.e., $\psi (x_1,x_2,t) = \psi (x_2,x_1,t)$, thus, $A(x_1,x_2,t)=A(x_2,x_1,t)$ and $S(x_1,x_2,t)=S(x_2,x_1,t)$. In the spin triplet state, on the other hand, we have $\psi (x_1,x_2,t) = - \psi (x_2,x_1,t)$, which is fulfilled either by $A(x_1,x_2,t)=-A(x_2,x_1,t)$ and $S(x_1,x_2,t)=S(x_2,x_1,t)$, or by $A(x_1,x_2,t)=A(x_2,x_1,t)$ and $S(x_1,x_2,t)=S(x_2,x_1,t)+\pi\hbar$.

{ Whichever rule we may adopt for $A(x_1,x_2,t)$ { and $S(x_1,x_2,t)$} under exchange, it follows from Eq.\ (\ref{eq:vel}) that  when the two electrons occupy the same position at a certain moment, they would also have the same velocity. { Furthermore, if follows from the symmetry of Eqs.\ (\ref{eq:qpot})-(\ref{eq:dyna}) under exchange of $x_1$ and $x_2$ that $x_1(t)=x_2(t)$ would hold all the time.} Therefore, we can conclude that the Bohmian { trajectories of} the two electrons do not cross each other.}

As is clear from the above discussion, this applies to not only fermions but also bosons. It should be stressed that the non-crossing law holds even when there is no classical interaction, such as the Coulomb repulsion, between the two identical particles. In such a case, crossing is prevented solely by the effect of exchange inherent in the quantum potential.}

{
Here, we illustrate this situation by a simple and striking example: two counter-propagating free particles. Let us suppose that the wave function of each particle is expressed as that of a Gaussian wave packet: 
\begin{eqnarray}
\label{eq:gaussian}
\psi^{g}(x,t) &\propto& \\ \nonumber
&\exp&[-\alpha(t) (x-x_{c}(t))^{2} +ip(x-x_{c}(t))+i\gamma(t)],
\end{eqnarray} 
with,
\begin{eqnarray}
\label{eq:alpha}
\alpha(t)&=& \frac{\alpha(0)}{1+2i\alpha(0)t},\\
\label{eq:xt}
x_{c}(t) &=& x_{c}(0)+pt,\\
\label{eq:gamma}
\gamma(t)&=& \frac{p^{2}t}{2}+\frac{i}{2} \ln \left[ 1+2i\alpha(0) t \right],
\end{eqnarray}
where $p$ denotes the momentum of the electron. By assuming a symmetrized spatial part corresponding to the singlet state of two electrons and no classical interaction between the two particles, then we can write the two-particle (spatial) wave function as,
\begin{eqnarray}
\label{eq:collision}
\psi_s(x_{1},x_{2},&t&) =\\
\frac{1}{\sqrt{2}}& & \left[ \psi^{g}_{1}(x_{1},t) \psi^{g}_{2}(x_{2},t) + \psi^{g}_{2}(x_{1},t)\psi^{g}_{1}(x_{2},t) \right].  \nonumber 
\end{eqnarray}
If we ignored (anti-)symmetrization, on the other hand, the wave function would read,
\begin{eqnarray}
\label{eq:collision-ns}
\psi_{ns}(x_{1},x_{2},t) = \psi^{g}_{1}(x_{1},t) \psi^{g}_{2}(x_{2},t).
\end{eqnarray}

In Fig.\ \ref{fig:nc} we plot the Bohmian trajectories corresponding to Eqs.\ (\ref{eq:collision}) and (\ref{eq:collision-ns}) for parameter values,
\begin{eqnarray}
\label{eq:coeffs}
\alpha(0)=0.2,  \quad p_{1} = -p_{2} = 3, \\
x_{c,1}(0) = - x_{c,2}(0) = -9,
\end{eqnarray}
and the initial positions,
\begin{eqnarray}
x_{1}(0)=x_{c,1}(0), \quad x_{2}(0) = x_{c,2}(0),
\end{eqnarray}
at the center of the packet. The velocities,
\begin{eqnarray}
\label{eq:vel-ec2ap}
v_{i}(t)= p_{i},
\end{eqnarray}
are constant for the case of Eqs.\ (\ref{eq:collision-ns}). For the case of Eq.\ (\ref{eq:collision}),
if further $\frac{x(t) \alpha(0)^{2} t}{p(1+4 \alpha(0)^{2}t^{2})} \ll 1$ holds, we can approximate the velocity as,
\begin{eqnarray}
\label{eq:vel-ec1ap}
&v_{i}&(x_{1},x_{2},t) =  \\ \nonumber
 & &\Re \left( \frac{p_{i}+p_{j}\exp[-8\alpha(t)x_{c,i}(t)x_{i}(t) + 4ix_{i}(t)p_{i}]}{1+\exp[-8\alpha(t)x_{c,i}(t)x_{i}(t) + 4ix_{i}(t)p_{i}]} \right),
\end{eqnarray}
%
\begin{figure}[t!]
\includegraphics[width=8cm, bb = 0 0 269 336]{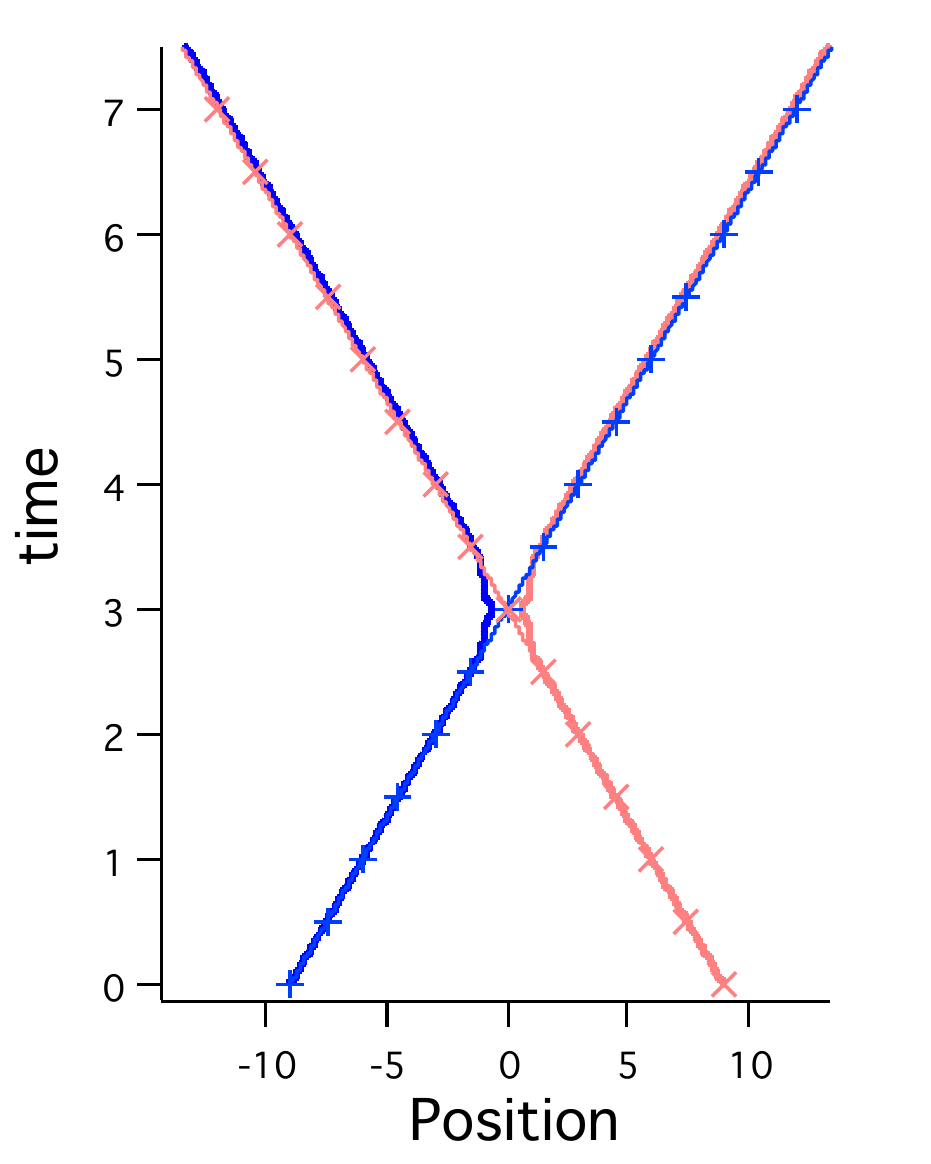}
\caption{{ }  Bohmian trajectories of two counter-propagating free particles, each expressed as a Gaussian wave packet. Bold lines and thin lines {with markers} correspond to Eqs.\ (\ref{eq:collision}) and (\ref{eq:collision-ns}), respectively (see text for details).}
\label{fig:nc}
\end{figure}
with $j$ being the other particle, which changes with time.
In Fig.\ \ref{fig:nc}, while the trajectories pass each other without being disturbed for the case of Eq.\ (\ref{eq:collision-ns}), they repel each other around the origin when we use Eq.\ (\ref{eq:collision}).It should be noticed that the latter [Eq.\ (\ref{eq:collision})] is the quantum-mechanically correct expression for identical particles. The significant repulsive behavior is seen when $\psi^{g}_{1}(x,t)$ and $\psi^{g}_{2}(x,t)$ overlap each other.

\newpage
\begin{figure}[t!]
\includegraphics[width=7.2cm, bb = 0 0 286 368]{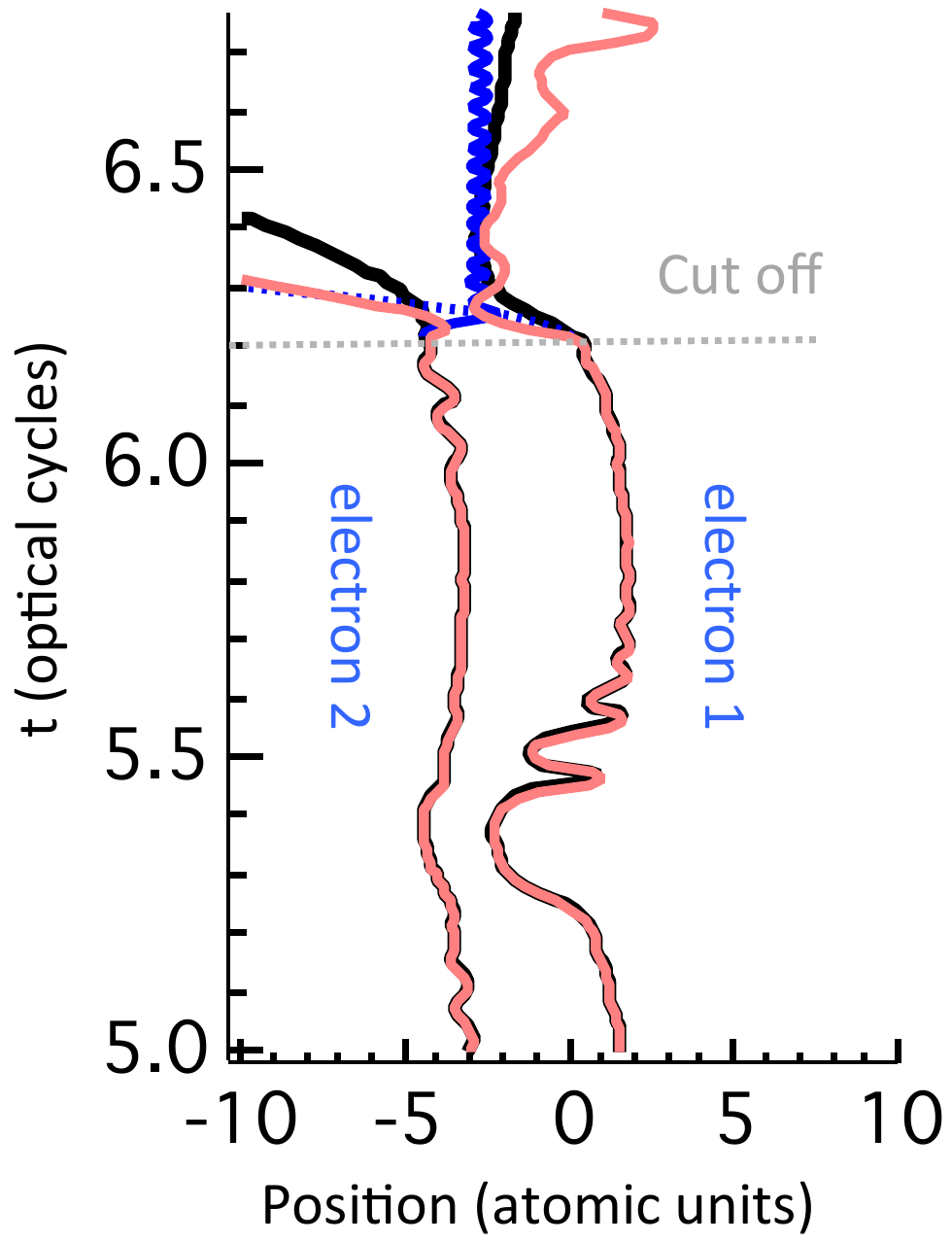}
\caption{{{ }  Comparison of trajectories with (black) both the Coulomb repulsion and quantum potential, without ({ pink}) the Coulomb repulsion after $t = 6.2T$, and without ({ blue}) the quantum potential after $t = 6.2T$. Black line is the same as in Fig.\ \ref{fig:types} (c), while in the case of the red and blue lines, the interelectronic Coulomb potential [the second term of Eq.\ (\ref{eq:ham})] and the quantum potential [second term of Eq.\ (\ref{eq:dyna})], respectively, are turned off at $t > 6.2T$ (gray dashed line).}}
\label{fig:aptr}
\end{figure}

In Fig.\ \ref{fig:aptr} we plot the Bohmian trajectories calculated for the same conditions as in Fig.\ \ref{fig:types} (c) but by switching off the interelectronic Coulomb repulsion [the second term of Eq.\ (\ref{eq:ham})] at $t > 6.2 T$ ({pink} line) { and by switching off the quantum potential [second term of Eq.\ (\ref{eq:dyna})] at $t > 6.2 T$ ({blue}line)}. One can indeed see electron 1 kick out electron 2 { after approaching faster even in the absence of the Coulomb repulsion ({pink} line). On the other hand, the Coulomb repulsion alone is not sufficient for the kick-out behavior ({blue line}). Rapid oscillation seen, e.g., around $5.5T$ is presumably a signature of a superposition of field-free stationary states, while that in the blue line at $t\gtrsim 6.3 T$ is due to classical oscillation around the left nucleus.}}

\bibliography{enhbib}

\end{document}